\documentclass{PoS}

\usepackage{amsmath}
\usepackage{amssymb}
\usepackage{amstext}
\usepackage{amsfonts}
\usepackage{graphicx}
\usepackage{subfigure}
\usepackage{multirow}
\usepackage{amssymb}
\usepackage{epsf}
\usepackage{epsfig}
\usepackage{epstopdf}
\usepackage[T1]{fontenc}

\newcommand{\ba}{\begin{eqnarray}}
\newcommand{\ea}{\end{eqnarray}}
\newcommand{\be}{\begin{equation}}
\newcommand{\ee}{\end{equation}}
\newcommand{\bd}{\begin{displaymath}}
\newcommand{\ed}{\end{displaymath}}
\newcommand{\een}{\nonumber\end{equation}}
\newcommand{\bea}{\begin{eqnarray}}
\newcommand{\eean}{\nonumber\end{eqnarray}}
\newcommand{\eea}{\end{eqnarray}}

\def\fig#1{Fig.~\ref{#1}}

\newcommand{\sla}{\displaystyle{\not}}

\newcommand{\beq}{\begin{equation}}   
\newcommand{\eeq}{\end{equation}}   
\newcommand{\beqn}{\begin{eqnarray}}  
\newcommand{\eeqn}{\end{eqnarray}}

\def\psibar{\overline{\psi}}

			\def\mcL{{\mathcal L}}

			\def\ba{{\bf a}}

			\def\bd{{\bf d}}

\newcommand{\old}[1]{}

\title{{\normalsize\hfill{\small CP3-Origins-2014-045
    DNRF90}}\\[15mm]{\vspace{-18mm} \normalsize\hfill{\small DIAS-2014-45 }}\\Scattering lengths in $SU(2)$ gauge theory with two 
  fundamental fermions}

\ShortTitle{Scattering lengths in a Composite Higgs Model}

\author{R. Arthur$\,^a$, \speaker{V. Drach}$^{a}$, M. Hansen$\,^a$, A. Hietanen$\,^a$,
  C. Pica$\,^a$, F. Sannino$\,^a$ \\
\llap{$^a$}{$CP^3$-Origins \& the  Danish Institute for Advanced Study  DIAS, University of Southern Denmark, Campusvej 55, DK-5230 Odense M, Denmark\\}
E-mail:  \email{drach@cp3.dias.sdu.dk}
}

\abstract{We investigate non perturbatively scattering properties of
  Goldstone Bosons  in an $SU(2)$ gauge theory with two Wilson fermions in the fundamental representation. Such a  theory can be used to build extensions of the Standard Model that unifies Technicolor and pseudo Goldstone composite Higgs models. The leading order contribution to the scattering amplitude of Goldstone bosons at low energy is given by the scattering lengths. In the context of technicolor extensions of the Standard Model the scattering lengths are constrained by WW scattering measurements.
We first describe our setup and in particular the expected chiral symmetry breaking pattern.
We then discuss how to compute them on the lattice and give preliminary results using finite size methods.}

\FullConference{The 32nd International Symposium on Lattice Field Theory,\\
		23-28 June, 2014\\
		Columbia University New York, NY}

\begin{document}

\section{Introduction}
% the model

In this work we consider an $SU(2)$ gauge field theory with two fermions
in the fundamental representation. The Lagrangian reads in the continuum :
\be\label{eq:lagrangian}
\mcL=-\frac{1}{4} F^a_{\mu \nu} F^{a~\mu\nu} + \psibar \left ( i  \sla D - m\right) \psi,
\ee
where $\psi=(u,d)$ is a doublet of Dirac spinor fields transforming according to the
fundamental representation.

 Because of the pseudo-realness of the fundamental representation of
 $SU(2)$, the mass term of the Lagrangian can be written in terms of 4
 Weyl spinors as follows\footnote{In fact $SU(2)$ is the first of the $Sp(2N)$ gauge theories. The associated conformal window was studied in \cite{Sannino:2009aw}. }
 \be
\mcL =-\frac{1}{4} F^a_{\mu \nu} F^{a~\mu\nu} + \psibar  i
  \sla D \psi + \frac{im}{2} Q^T (-i \sigma_2) CE Q  +  \left(Q^T (-i \sigma_2) CE Q)\right)^\dagger
\ee
where $\sigma_2$ acts on color indices and  $C$ is the
charge conjugation  matrix. Furthermore, we have defined :

\be
Q = \begin{pmatrix} u_L \\ d_L \\ -i\sigma_2 C \bar{u}_R^T \\
  -i\sigma_2
  C\bar{d}_R^T\end{pmatrix},\text{and}\quad E=\begin{pmatrix}0 &0
   & +1 & 0\\
0  &0  & 0 & +1\\
-1 &0  & 0 & 0\\
0  &-1 & 0 & 0\\
 \end{pmatrix}.
\ee
We have used $q_{L,R} = P_{L,R}
q,\bar{q}_{L,R} = \bar{q} P_{R,L}$ with  $P_L=\frac{1}{2}
(1-\gamma_5)$ and $P_R=\frac{1}{2} (1+\gamma_5)$.
The model  exhibits an $SU(4)$ flavour symmetry in the massless
limit. To fix notations, the 15 generators of the corresponding Lie
algebra will be denoted  $T^{a=1,\dots,15}$. 
After adding a mass term, the remnant flavour symmetry is the
group spanned by the algebra that preserves $E T^{a,T} + T^{a,T} E = 0$.  This
relation defines the  10-dimensional algebra of the $SP(4)$ group.
The chiral symmetry breaking pattern is thus expected to be $SU(4)
\rightarrow SP(4)$ leading to 5 Goldstone bosons.

% general motivations for studying scattering properties
As proposed in \cite{Cacciapaglia:2014uja}, the Lagrangian Eq.~(\ref{eq:lagrangian}) can be
embedded into the Standard Model in such a way that it interpolates
between composite Goldstone Higgs and Technicolor models\cite{Ryttov:2008xe,Appelquist:1999dq}.  

The model has been investigated on the lattice in \cite{Hietanen:2014xca}, and  the
chiral symmetry breaking pattern  has been proven to be the expected
one \cite{Lewis:2011zb}. Updated results concerning our on-going
effort are summarized in \cite{Arthur:2014lma}.

% decay width / constraint from LHC on WW scattering 
Since one feature of the model is to  provide a dynamical explanation
of Electroweak symmetry breaking, calculating the scattering properties of the Goldstone bosons of the
underlying theory can be related to scattering properties of
longitudinal W bosons according to the equivalence  theorem\cite{Lee:1977eg}. 

% Classification of the two GB states

The two particle states involving two Goldstone bosons can  be
classified according $ \bf{5} \otimes \bf{5} = \bf{1} \oplus \bf{10} \oplus \bf{14}$ and it can be
shown that $\pi^+ \pi^+$ belongs to the $\bf{14}$ dimensional representation
of $SP(4)$. 

% LO theorem 

The low energy prediction for this realization of chiral
symmetry breaking has been considered in\cite{Bijnens:2011fm}, and reads:
\be\label{eq:LO}
m_{\rm{PS}} a^{\bf{14}}_{0,LO} = - \frac{m_{\rm{PS}}^2}{32 f_{\rm{PS}}^2}
\ee
where $m_{\rm{PS}}$ and $f_{\rm{PS}}$ are respectively the mass and
decay constant of the Goldstone bosons, and where $a^{\bf{14}}_{0,LO} $ is
the scattering length related to the
partial wave amplitude at zero-momentum transfer of the scattering process : $\pi^+\pi^+ \rightarrow \pi^+
\pi^+$. The superscript $\bf{14}$ refers to the dimension of the
corresponding irreducible representation
of $SP(4)$. 

\section{Lattice techniques}

% Luscher formula 
% ADD DISCussion of higher order term
% add discussion of finite T
% Fierz rearramgement
% number of stoch. samples

In order to obtain the scattering lengths from lattice simulations in Euclidean
space, we used the same strategy as in QCD to compute scattering
lengths of pions in the isospin 2 channel (see \textit{e.g} \cite{Feng:2009ij}). We briefly review here our approach.

The general idea is to compute the energy of the two Goldstone bosons
in a finite box. L\"uscher's finite size method \cite{Luscher:1986pf}, relates the energy shift due to the
interactions of the two particles in a finite box  to
their infinite volume scattering length according to the following
formula:
\be\label{eq:FSM}
\frac{\delta E_{\pi \pi}}{m_\pi}  =\frac{4\pi m_{\rm{PS}} a^{\bf{14}}_{0}} { (m_{\rm{PS}}L)^3} \left[1 + c_1 \frac{ m_{\rm{PS}} a^{\bf{14}}_{0}}{m_{\rm{PS}} L}  + c_2 \left(\frac{ m_{\rm{PS}} a^{\bf{14}}_{0}}{m_{\rm{PS}} L}\right)^2 \right],
\ee
where $c_{1,2}$ are two known constants, $L$ is the spatial extension of
the lattice and $m_{\rm{PS}}$ is the Goldstone boson mass.

\begin{figure}[h!] % correct box .
\includegraphics[width=\textwidth]{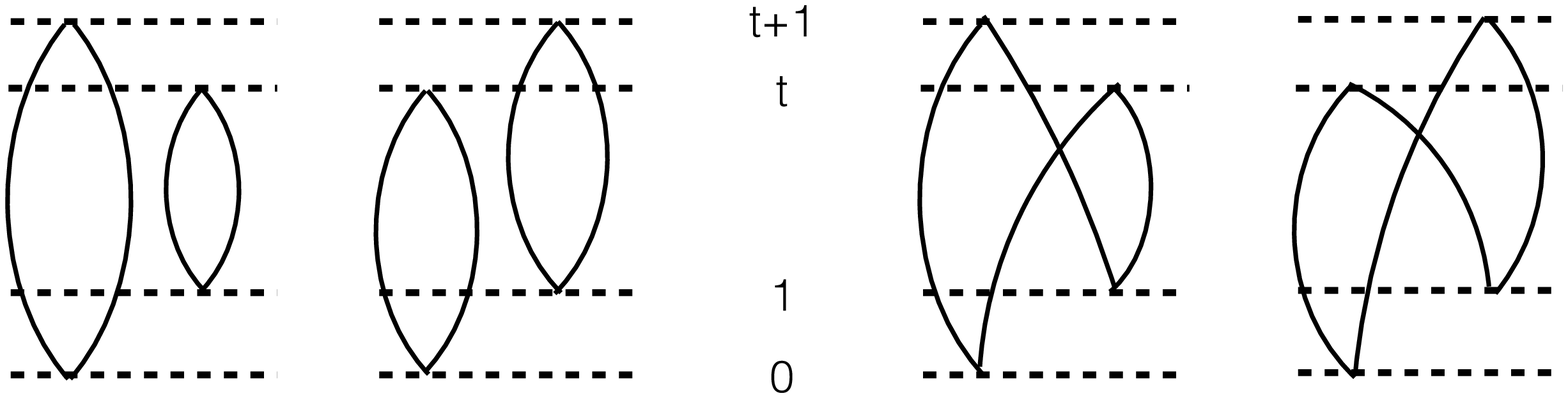}
\caption{Illustration of the Wick contractions.}\label{fig:wick}
\end{figure}

The interpolating fields for the single and two-particle operator are defined as follows:
% Operators
\be
 \pi^+(t)  \equiv  \sum_{\vec{x}}  \bar{d}\gamma_5  u (\vec{x},t),
 \qquad (\pi^{+}\pi^{+})(t) \equiv \pi^{+}(t+a) \pi^{+}(t) ~,
\ee
where $a$ denotes the lattice spacing. Note that the operator for the
two-particle state is defined using two single particle states at
different time slices in order to avoid issues with Fierz
rearrangement \cite{Fukugita:1994ve}. The following two-point functions can then be built:
% Correlators
\begin{eqnarray}
C_{\pi}(t) &=& \langle (\pi^+)^\dagger(t+t_s)  \pi^+(t_s) \rangle \\
C_{\pi\pi}(t) &=&\langle (\pi^+ \pi^+)^\dagger(t+t_s)
(\pi^+\pi^+)(t_s) \rangle
\end{eqnarray}
where $t_s$ is the source time slice.
% Contractions
The Wick contractions are illustrated in Fig. \ref{fig:wick}. In order
to estimate the necessary correlators, we used stochastic estimators
with $Z_2$ noise on a single time slice.
% Improved ratio and asymptotic behaviour
The ratio 
\be\label{eq:R}
R(t+a/2) \equiv \frac{C_{\pi\pi}(t) -C_{\pi\pi}(t+a) }{C^2_{\pi}(t) -C^2_{\pi}(t+a)}
\ee
can be shown to have the following asymptotic behavior for large  $t$ :
\be\label{eq:R_asymptotic}
R(t+a/2) = A_R \left[\cosh(\delta E_{\pi \pi} t') + \sinh(\delta
  E_{\pi \pi} t') \coth( 2m_{\rm{PS}} t') \right]
\ee
with $t'= t+a/2-T/2$.
This ratio, designed to remove a constant contribution due the
finite extension in time $T$, is sometimes referred as the derivative
method \cite{Umeda:2007hy}.

\section{Results}

\begin{figure}[h] % correct box .
\centering
\includegraphics[width=0.8\textwidth]{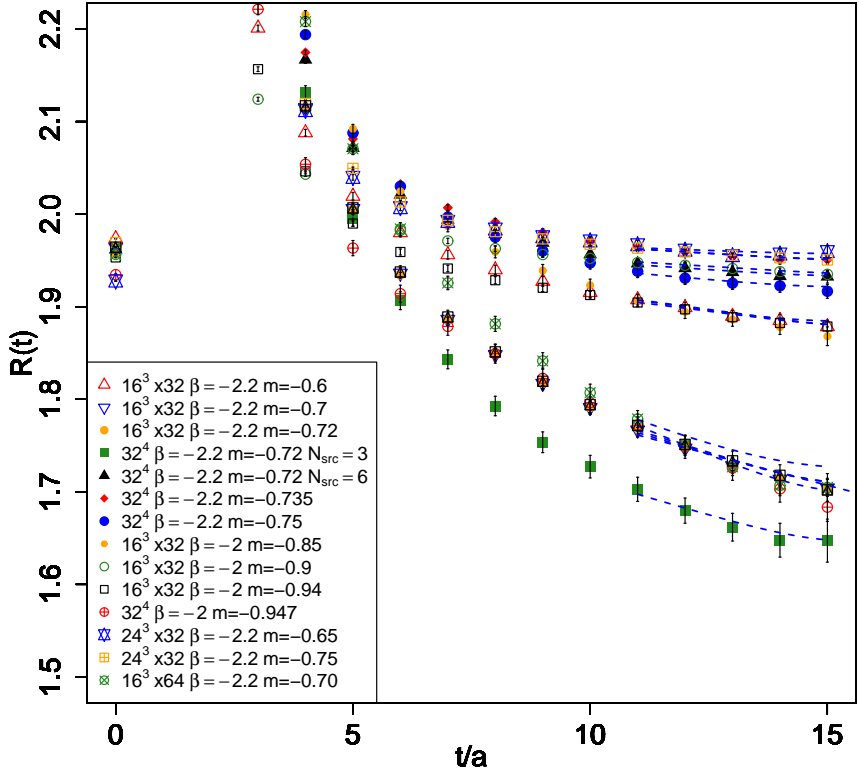}
\caption{Ratio $R$ defined in Eq.~(\protect\ref{eq:R}) as a function of time $t/a$ for our various gauge ensembles with different spatial
volumes, lattice time extent, lattice spacing and fermion
masses. The blue dotted curves illustrate fits according to Eq.~(\protect\ref{eq:R_asymptotic}).}\label{fig:ratio_sym}
\end{figure}

 %raw results and  fitting window
We show in \fig{fig:ratio_sym} the ratio $R$ as a function of time $t/a$ for our various gauge ensembles with different spatial
volumes, lattice time extent, lattice spacing and fermion masses.  As can
be seen the ratio $R(t)$ is determined accurately in all our
simulations. We also illustrate a particular choice of fitting range
by a blue dotted curve.

In order to extract the scattering length $a_0^{\bf{14}} m_{\rm{PS}}$
we can then perform a two parameter fit  of the correlator on a given
range $[t_1,T/2]$. Using
Eq. (\ref{eq:FSM}) one can then obtain an estimate of $a_0^{\bf{14}}
m_{\rm{PS}}$. Note that $m_{\rm{PS}}$ has been determined in
\cite{Hietanen:2014xca} using standard techniques.

\begin{figure}[h] % correct box .
\centering
\includegraphics[width=0.8\textwidth]{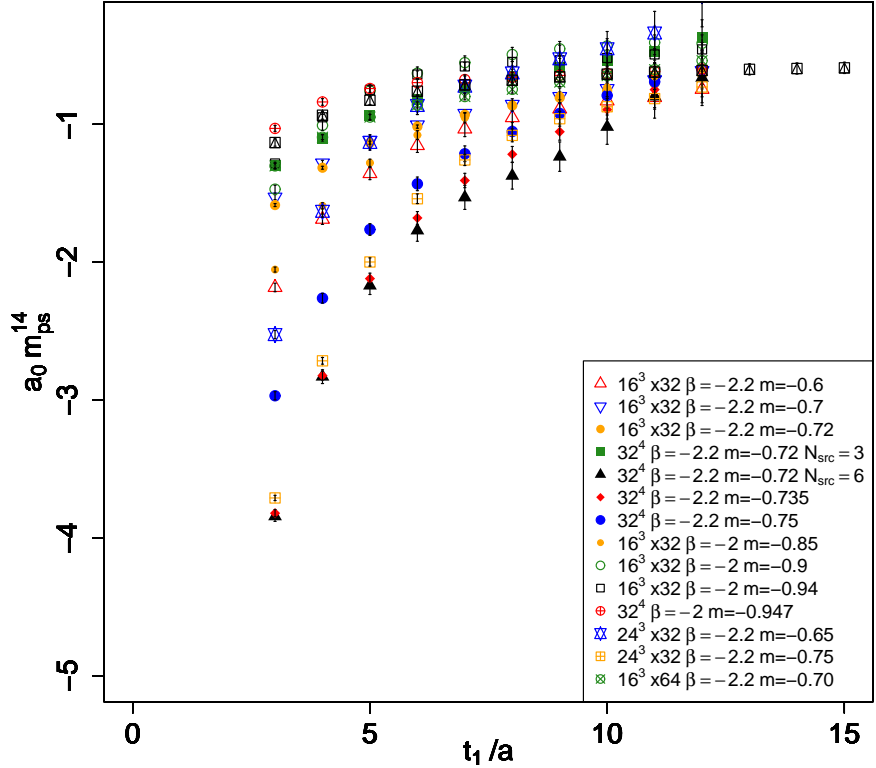}
\caption{ Dependence of the scattering length
extracted from the $R(t)$ as a function of the lower bound of
the fitting window $t_1$. Results show a significant dependence,
indicating a large excited states contribution.}\label{fig:a_vs_t1}
\end{figure}

In \fig{fig:a_vs_t1} we show the dependence of the scattering length
extracted from the previous figure as a function of the lower bound of
the fitting window $t_1$. For all the gauge ensembles considered in
this work, results show a pronounced dependence
indicating a large excited states contamination. This may be due to
the use of a local (\textit{i.e} not smeared) operator. However for
$t_1\ge 11a $ one observe that the results do not depend on the fitting
range indicating that the asymptotic regime is reached. In what
follows, we thus obtain the value of the scattering length
choosing $t_1=11a$ on all our ensembles.

Note that we also checked the convergence of the expansion Eq. (\ref{eq:FSM}),
by looking a posteriori that the relative contribution of the term
proportional to $c_1$ and $c_2$ are small. In practice we
observe that the term proportional to $c_1$ provides $\sim 10\%$ of $\delta
E_{\pi\pi}/m_{\rm{PS}}$ while the term  proportional to $c_2$ contribute to
   the order $\sim 1\%$. It is thus reasonable to assume that
 Eq. \ref{eq:FSM} is valid and that one does not need to consider higher
 order corrections in $\frac{m_{\rm{PS}} a_0^{\bf{14}}}{m_{\rm{PS}} L}$
 for all our ensembles. 

% finite T

\begin{figure}[h] % correct box .
\centering
\includegraphics[width=0.8\textwidth]{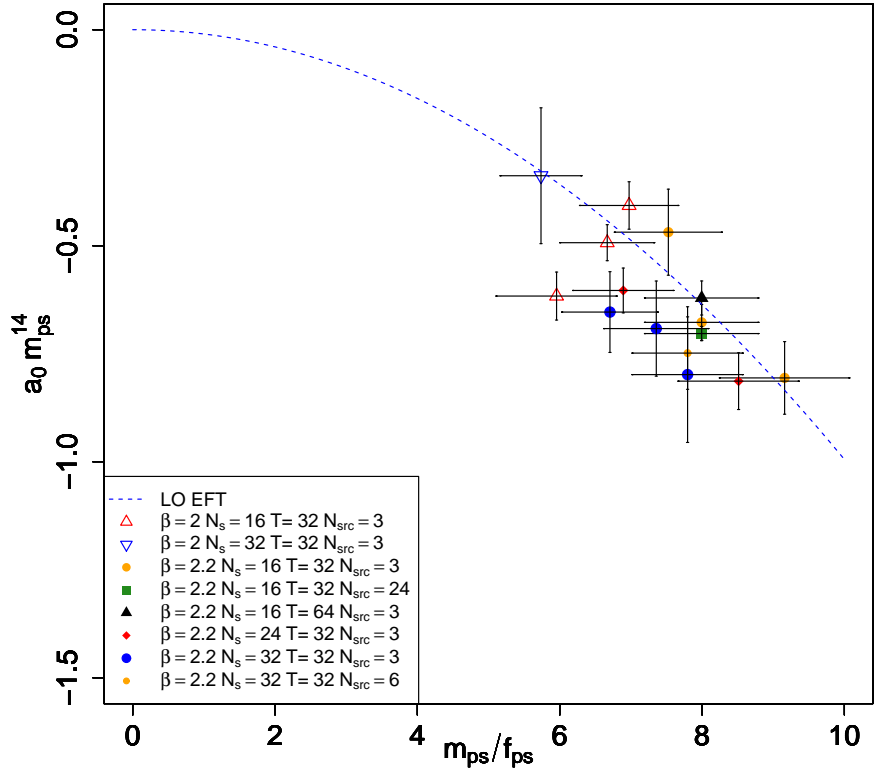}
\caption{$m_{\rm{PS}}a^{\bf{14}}_0$ as a function of the dimensionless ratio
$m_{\rm{PS}}/f_{\rm{PS}}$ for all our ensemble. The vertical error bar
are statistical and the horizontal one, comes mostly from the
uncertainty on the perturbative value of the renormalization constant of the pseudoscalar meson
decay constant. The blue dotted line shows the leading order prediction
from  Eq.~\protect\ref{eq:LO}. The parameters $\beta$,$N_s$ (spatial
extent), $T$ (time extent) and
$N_{\rm{src}}$(number of stochastic source per configuration) are indicated in the legend.}\label{fig:chiral}
\end{figure}

\newpage                          
% Chiral behaviour
In \fig{fig:chiral} we summarize our findings by showing the value of
$a^{\bf{14}}_0 m_{\rm{PS}}$ as a function of the dimensionless ratio
$m_{\rm{PS}}/f_{\rm{PS}}$ for all our ensembles. Note that
$f_{\rm{PS}}$ is renormalized perturbatively. Conservatively we thus
choose a $20\%$ error bar on the value of the renormalized
pseudoscalar decay constant which dominates the horizontal error
bar. Vertical error bar are purely statistical and in particular they do not take into
account systematic error on the particular choice of a plateau
range. However as we argued previously the systematic is most likely
to be small compare to our statistical error in view of our precedent discussion.We also show the LO
prediction from effective field theory as in Eq.~\ref{eq:LO}~\cite{Bijnens:2011fm}. As mentioned in the introduction the LO prediction do not
depend on any low energy constants. Within our current statistical and
systematic errors the lattice results are compatible with the LO order predictions.

% FV effects

\newpage
\section{Conclusion}

In this study we successfully applied finite size method to the two
Goldstone bosons system in an $SU(2)$ gauge theory with two fundamental
fermions. We argued that the determination of scattering lengths and
more precisely of the corresponding low energy constant can be used to
constrain the WW coupling.
The result in the particular channel we consider shows good agreement
with the leading order prediction from effective field theory. 
We plan to extend this work to the study of the other channels and in
particular to the determination of vector meson decay width which is
of fundamental relevance for LHC phenomenology.

%\vspace*{-0.5cm}
\section*{Acknowledgments}
%\vspace*{-0.5cm}

This work was supported by the Danish National Research Foundation
DNRF:90 grant and by a Lundbeck Foundation Fellowship grant. The computing facilities were provided by the Danish Centre for Scientific Computing.

%\vspace*{-0.5cm}

\end{document}